\documentclass[preprint,superscriptaddress,12pt]{elsarticle}
\usepackage{}
\usepackage{bbm}

\usepackage{amsmath,bm}% bold math
\usepackage{graphicx}% Include figure files
\usepackage{dcolumn}% Align table columns on decimal point
\usepackage{tipa}
\usepackage{enumerate}
\usepackage{amsfonts}
\usepackage{graphicx}
\usepackage{amssymb}
\usepackage{cases}
\usepackage{multirow}
\usepackage{booktabs}
\usepackage{appendix}
\usepackage{color}

\newtheorem{theorem}{Theorem}
\newtheorem{corollary}{Corollary}
\newcommand{\proof}[1]{{\bf Proof.} #1 $\Box$.}
\newcommand{\tabincell}[2]{\begin{tabular}{@{}#1@{}}#2\end{tabular}}
\allowdisplaybreaks

\journal{Theoretical Computer Science}

\begin{document}

\begin{frontmatter}

\title{Security Improvements of Several Basic Quantum Private Query Protocols with O(log N) Communication Complexity}

\author[rvt,Cze]{Fang Yu}
\author[focal,Cze,Por]{Daowen Qiu \corref{cor1}}
\author[rvt]{Xiaoming Wang}%
\author[els]{Qin Li}
\author[focal]{Lvzhou Li}%
\author[Cze]{Jozef Gruska}

\cortext[cor1]{yufangh4@jnu.edu.cn (Fang Yu), issqdw@mail.sysu.edu.cn (Daowen Qiu)}

\address[rvt]{Department of Computer Science and Technology, Jinan University. Guangzhou, China}
\address[focal]{Institute of Computer Science Theory, School of Data and Computer Science, Sun Yat-sen University. Guangzhou, China}
\address[els]{College of Information Engineering, Xiangtan University. Xiangtan, China}
\address[Cze]{Faculty of Informatics, Masaryk University, Brno, Czech Republic}
\address[Por]{ Instituto de Telecomunica\c{c}\~{o}es, Departamento de Matem\'{a}tica, Instituto Superior T\'{e}cnico, Universidade de Lisboa, Av. Rovisco Pais 1049-001, Lisbon, Portugal}

\begin{abstract}
New quantum private database (with $N$ elements) query protocols are presented and analyzed. Protocols preserve $\mathcal{O}(\log N)$ communication complexity of known protocols for the same task, but achieve several significant improvements in security, especially concerning user privacy. For example, the randomized form of our protocol has a cheat-sensitive property - it  allows the user to detect a dishonest database with a nonzero probability, while the phase-encoded private query protocols \cite{Ole12,Yu14} for the same task do not have such a property. Moreover, when the database performs the computational basis measurement, a particular projective measurement which can cause a significant loss of user privacy in the previous private query protocols with $\mathcal{O}(\log N)$ communication complexity, at most half of the user privacy could leak to such a database in our protocol, while in the QPQ protocol \cite{GLM08}, the entire user privacy could leak out. In addition, it is proved here that for large $N$, the user could detect a cheating via the computational basis measurement, with a probability close to $\frac{1}{2}$ using $\mathcal{O}(\sqrt{N})$ special queries. Finally, it is shown here, for both forms of our protocol, basic and randomized, how a dishonest database has to act in case it could not learn user's queries.
\end{abstract}


\end{frontmatter}

\section{Introduction}
The symmetrical private information retrieval (SPIR) problem \cite{Chor95,GIKM00} is usually modeled by a user querying a database $A$, actually one of its $N$ items, say $A_{j}$ (what is usually a bit and  $j \in [N]$), while keeping private not only the value of $j$ (so-called \textit{user privacy}), but also all other items $A_{k} (k\neq j)$ (so-called \textit{data privacy}). This is one of the fundamental problems in the area of secure multiparty computation and communication. In the classical world, solutions to this problem rely on unproven computational hardness assumptions from the complexity theory, in order to guarantee privacy of both parties involved \cite{Chor95}. Actually, no perfect solution to this problem seems to be known even in the quantum world \cite{Lo97,KW03}.

On the other hand, for this fundamental problem, significant advantages in communication efficiency can be obtained using quantum resources. Only $\mathcal{O}(\log N)$ of data is needed to be exchanged in some of the known protocols \cite{GLM08,Ole12,Yu14}, which allows an exponential reduction in the communication complexity when comparing with the best classical SPIR protocols proposed so far \cite{Chor95,BIKR02,Ambainis97}. While these protocols can achieve their querying goals with the lowest possible level of communication complexity (considering that $j$ has a coding of $\log N$ length), that is with $\mathcal{O}(\log N)$ communication complexity, these protocols are secure only when the parties are honest \cite{Bro15}. A dishonest database may acquire a significant amount of information about queries, via easy-to-implement such an operation as a single-qubit computational basis measurement (with two basis states $|0\rangle$ and $|1\rangle$). Therefore, any improvement on security of such protocols, especially concerning the user privacy, is an interesting and important challenge.

The QPQ protocol \cite{GLM08} has been the first private query protocol with $\mathcal{O}(\log N)$ communication complexity. It uses a qRAM algorithm \cite{GLM08qram} to provide answers to queries. The other two private query protocols with $\mathcal{O}(\log N)$ communication complexity, namely \cite{Ole12,Yu14}, complete the query task via a method which encodes item's values in the phase of the transmitted state. This encoding is performed either by an oracle \cite{Ole12}, which is assumed to be able to recognize solutions to the search problem, or by a unitary operator the form of which is given explicitly \cite{Yu14}.

All three protocols \cite{GLM08,Ole12,Yu14} rely on setting a unique, so-called rhetoric, query \footnote{In \cite{GLM08}, the query $0$ is chosen to be rhetoric if it has a known standard answer $A_0$, say $0$. In our protocols, there is no need for a rhetoric query to have such a standard answer. That is to say, any query different from the true query $j$ can serve as a rhetoric query. } to secure the true query $j$. The way is to let the user always prepare his/her queries by coherently mixing the true query $j$ with the rhetoric query. Due to such a superposition the database can not learn $j$ immediately but needs to perform some cheating operations on the state in order to obtain information. However, the operations can cause deviation on the state and thus could be detected by the user after he/she received the state returned from the database. In terms of this approach, the QPQ protocol \cite{GLM08} can obtain a cheat-sensitive property, i.e. the user can have a non-zero probability to detect dishonesty of the  database. (Cheat sensitive cryptographic protocols between mistrustful parties, are in literature \cite{HK04}  defined as protocols which guarantee that, when either of them cheats, the other has a nonzero probability to detect such a cheating.) In addition, setting a rhetoric query with a standard answer can guarantee the success of the query (by getting the correct $A_j$ value) not only in the QPQ protocol \cite{GLM08}, but also in phase-encoded private query protocols \cite{Ole12,Yu14}. Therefore, such a setting is essential for both protocols.

Both protocols have significant deficiencies concerning security, even in the case of an attack via a simple computational basis measurement. The QPQ protocol will inevitably leak $j$ to a dishonest database performing such an attack, despite of the fact that it would be cheat-sensitive. In contrast, the phase-encoded private query protocols \cite{Ole12,Yu14} could leak at most half amount of the user privacy to such a database, but they are not cheat-sensitive. That means that such a database can prevent exposure to detections, even if the attack results in a significant information leakage. Such an attack can even ruin the protocol because it can lead to an unsuccessful query (by returning only a random answer). Papers \cite{Ole12,Yu14} propose variants of the above protocols that aim to enhance security. However, none of them can deal successfully with such easy-to-implement attacks.

In this paper, we propose a new scheme for private query protocols which is also of $\mathcal{O}(\log N)$ communication complexity. In this scheme, the database retrieves requested items using a special data retrieving algorithm that is running locally at the side of the database, for example, the qRAM algorithm used in the QPQ protocol \cite{GLM08}, and the user extracts the target item using a method which is developed from one that was used in the phase-encoded private query protocols \cite{Ole12,Yu14}. This method allows the user to make a superposition of the true query with several rhetoric queries, without having answers to these rhetoric queries known to the user beforehand. Despite lacking a preliminary knowledge about answers to rhetoric queries (or even rhetoric queries themselves in a randomized variant of the protocol), the user can still get the correct answer to his/her real query.

A use of multiple rhetoric queries can enhance security concerning user privacy in private query protocols. In the basic form of the proposed protocol, all queries but $j$ can be rhetoric queries, which is an assumption known to the database. This form of our new protocol is not cheat-sensitive, but can be easily modified to a variant which has a cheat-sensitive property. In such a variant, the user is allowed to select rhetoric queries randomly. This makes it difficult for the database to conceal its cheating activities due to the lack of knowledge about rhetoric queries and hence enhance security in preserving user privacy in a phase-encoded private query protocol. In addition, the setting of multiple rhetoric queries may lead to a much smaller amount of information leakage about the query in private query protocols that use a special data retrieving algorithm to provide answers to queries. Therefore, the proposed schemes can be regarded as generalizations of known private query protocols with $\mathcal{O}(\log N)$ communication complexity in the sense that they not only maintain their advantages in communication complexity, but also achieve an improvement in security, especially in preserving user privacy.

The rest of the paper is organized as follows. Section \ref{Sec_protocol} starts with a brief review of related papers. A detailed description of the basic form of our protocol, as well as an illustration of the protocol actions by a figure is given in this section. In Section \ref{Sec_security}, the security of our new protocol is analyzed. After that, a variant of our new protocol is presented in Section \ref{Sec_variant} in which the number of rhetoric queries is introduced as a variable. Security improvements are then analyzed with respect to that variable. In particular, the case of an easy-to-implement, but destructive, computational basis measurement through which the database steals user privacy is addressed in details. Conclusions are given in Section \ref{Sec_conclusion}.

\section{Related Papers and New Protocol}\label{Sec_protocol}

\subsection{Related Papers}
In the QPQ protocol \cite{GLM08}, the user prepares at the beginning two states $|j\rangle$ and $\frac{1}{\sqrt{2}}(|j\rangle+|0\rangle)$, where $\{|i\rangle, i \in [N]\}$ is the computational basis of the quantum space the user works within. The user then sends randomly one of the two states to the database and waits for a response from the database before sending the next one. The database runs the qRAM algorithm \cite{GLM08qram}, with two prepared states, returning as the output states $|j\rangle|A_j\rangle$ and $\frac{1}{\sqrt{2}}(|j\rangle|A_j\rangle+|0\rangle|A_0\rangle)$, respectively. The user gets $A_j$ by measuring the state $|j\rangle|A_j\rangle$ in the computational basis and then uses $A_j$ with the standard value of $A_0$ to construct a measurement to test whether the second state is in the expected form $\frac{1}{\sqrt{2}}(|j\rangle|A_j\rangle+|0\rangle|A_0\rangle)$.

In one of the phase-encoded private query protocols from \cite{Ole12}, the user prepares initially the state $\frac{1}{\sqrt{2}}(|j\rangle+|0\rangle)$. The database is then
expected to return the state $\frac{1}{\sqrt{2}}\left((-1)^{A_j}|j\rangle+(-1)^{A_0}|0\rangle\right)$, with the answer being encoded in the phase of the query element. The encoding process is realized by an oracle, which is assumed to be able to recognize the correct solutions for the search problem. The user then can get the value of $A_j \oplus A_0$ by discriminating $\frac{1}{\sqrt{2}}(\pm|j\rangle+|0\rangle)$, and can then figure out the value of $A_j$ with respect to the standard $A_0$. The entangled state $\frac{1}{\sqrt{2}}(|j,0\rangle+|0,j\rangle)$ is actually used to enhance security in the other phase-encoded private query protocol \cite{Yu14}.

From the above descriptions we can see that in both protocols, only one rhetoric query, namely ``0", is used. It masks the true query $j$ so that the private $j$ is unlikely to leak to an honest database and can only be learned by a dishonest database in a non-deterministic manner. The fact that the rhetoric query $0$ has a standard answer $A_0$ guarantees not only a successful query (by extracting the correct answer $A_j$ to the query $j$) in the phase-encoded private query protocols, but also the cheat-sensitivity of the QPQ protocol.

However, it is important to observe that using a rhetoric query known to the database may cause serious security problem in the above private query protocols with $\mathcal{O}(\log N)$ communication complexity. Indeed, when measuring the received state in the computational basis, the database can learn $j$ (in case of a nonzero outcome) with high probability, which is $1$ in the case of the QPQ protocol and $1/2$ for the phase-encoded private query protocols. Moreover, in the phase-encoded private query protocols, the database can conceal itself easily by merely sending back the measurement outcome states $|j\rangle$ or $|0\rangle$, because no unexpected outcome could be measured later by the user. This suggests that the database can make a large  gain at a low cost in the known protocols by means of such an action.

In terms of the data privacy preservation, the phase-encoded private query protocols exhibit better performance than the QPQ protocol. A dishonest user can learn at most $1$ bit of information from the database in the phase-encoded private query protocols whilst in the QPQ protocol, by making use of one extra round of the communication designed specifically for the honesty detection, user can obtain one additional bit.

\subsection{Our New Protocol - Basic Form}
The user starts the protocol by preparing the state $|\psi_0\rangle=\frac{1}{\sqrt{2}}\left(|j\rangle+\frac{1}{\sqrt{N-1}} \sum_{j'\neq j}|j'\rangle\right)$ and then sends it to the database. In order to store answers to queries, the database prepares an answering qubit initially in the state $|0\rangle$. After receiving the state $|\psi_0\rangle$, the database applies the special data retrieving algorithm to convert $|\psi_0\rangle|0\rangle$ into
\begin{quote}
\centering
$|\Psi_1\rangle=\frac{1}{\sqrt{2}}\left(|j\rangle|A_j\rangle+\frac{1}{\sqrt{N-1}}\sum_{j'\neq j}|j'\rangle|A_{j'}\rangle\right)$.
\end{quote}

\noindent As the next step, the database sends the state $|\Psi_1\rangle$ back to the user, who then applies on it the controlled addition modulo 2 operation $^\wedge\oplus$ with one ancillary qubit being initially in the state $|q\rangle=\frac{1}{\sqrt{2}}(|0\rangle-|1\rangle)$. The corresponding action of the operation is
\begin{equation*}
|j\rangle|A_j\rangle|q\rangle \rightarrow |j\rangle|A_j\rangle|q\oplus A_j\rangle, \ \ \ \ \ \
|j'\rangle|A_{j'}\rangle|q\rangle \rightarrow |j'\rangle|A_{j'}\rangle|q\rangle,
\end{equation*}

\noindent i.e. $|q\rangle$ is flipped when control qubits are in the state $|j\rangle|1\rangle$, and remains unchanged otherwise. Afterwards the state $|\Psi_1\rangle|q\rangle$ evolves into the state\
\begin{small}
\begin{eqnarray*}
^\wedge\oplus(|\Psi_1\rangle|q\rangle) &=& \frac{1}{\sqrt{2}}\left(|j\rangle|A_j\rangle|q\oplus A_j\rangle+\frac{1}{\sqrt{N-1}}\sum_{j'\neq j}|j'\rangle|A_{j'}\rangle|q\rangle\right) \\
&=& \frac{1}{\sqrt{2}}\left((-1)^{A_j}|j\rangle|A_j\rangle+\frac{1}{\sqrt{N-1}}\sum_{j'\neq j}|j'\rangle|A_{j'}\rangle\right)\otimes |q\rangle=|\Psi_2\rangle|q\rangle.\label{Eq_phi2}
\end{eqnarray*}
\end{small}

\noindent As the next step, the user keeps the state $|q\rangle$ and sends $|\Psi_2\rangle$ back to the database, which reverses the state of the answering qubit into the state $|0\rangle$ by querying the data retrieving algorithm again. Then the state $|\Psi_2\rangle$ evolves into the state
\begin{small}$$|\Psi_3\rangle = \frac{1}{\sqrt{2}}\left((-1)^{A_j}|j\rangle+\frac{1}{\sqrt{N-1}}\sum\limits_{j'\neq j}|j'\rangle\right)\otimes|0\rangle = |\psi_3\rangle|0\rangle, \label{Eq_phi3}$$ \end{small}

\noindent For convenience, we denote $\frac{1}{\sqrt{2}}(|j\rangle+\frac{1}{\sqrt{N-1}}\sum_{j'\neq j}|j'\rangle)$ by $|\psi_3^+\rangle$ and $\frac{1}{\sqrt{2}}(-|j\rangle+\frac{1}{\sqrt{N-1}}\sum_{j'\neq j}|j'\rangle)$ by $|\psi_3^-\rangle$, corresponding to the $0$ or $1$ value of $A_j$, respectively. Note also that the states $|\psi_3^\pm\rangle$ are orthogonal. Therefore, the user can perform a discriminating measurement after receiving a state from the database, which keeps the answering qubit, and then interprets that
\begin{equation}
\left\{
\begin{array}{l@{\ }l}
A_{j}= 0, & \textrm{if} \ |\psi_3^+\rangle \ is \ measured, \\
A_{j}= 1, & \textrm{if} \ |\psi_3^-\rangle \ is \ measured, \\
the \ database \ is \ cheating,& \textrm{if} \ neither \ of \ |\psi_3^\pm\rangle \ is \ measured.
\end{array} \\
\right.
\label{Eq_distinguish}
\end{equation}

The whole procedure is briefly illustrated in Fig.1.
\begin{figure}[t]
\centering
\includegraphics[width=8.2cm]{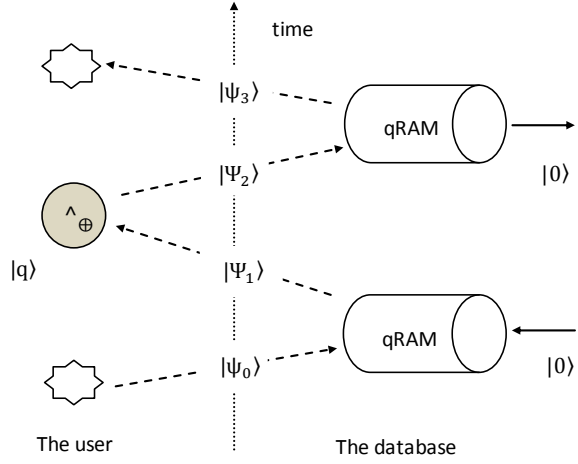}
\caption{The protocol. ( $\hat{}_{\oplus}$ is the control addition modulo 2 operator.)}
\label{fig:figure1}
\end{figure}

%Tu pokracovat

In the above protocol, two rounds of communications are needed to get an answer to a single query. During such communications, $\log N$ qubits, which store queries, are transmitted four times. In addition, $1$ qubit for storing answers to queries is transmitted twice. In total, $4\log N+2$ qubits need to be transmitted, i.e. the communication complexity of the protocol is asymptotically $\mathcal{O}(\log N)$, which is an exponential reduction comparing with the classical and quantum private query protocols known so far. Note that the data retrieving algorithm runs simply locally at the side of the database. Therefore, its time cost has no contribution for the computational complexity.

In the protocol, all item's indexes, except $j$, are used as rhetoric queries to create the initial superposed state. From the above procedure we can see that the user will obtain the correct answer $A_j$ to his/her query $j$ deterministically, without having to know answers to rhetoric queries beforehand. Moreover, the setting of the multiple rhetoric queries provides better security, especially a better user privacy, than previously mentioned protocols. Security of the protocol, with respect to such a setting, is considered in the following two sections.

\section{Security Issues} \label{Sec_security}
Similarly to all already known communication-efficient private query protocols, our new protocol is secure with respect to honest parties. Indeed, on one side, data privacy can be seen as being preserved because there is only one database item retrieved by a user that follows protocol steps. On the other side, due to the masking effect of rhetoric queries an honest database can not learn the true query $j$ from the received superposed states. Therefore, the user privacy is preserved when honest adversaries communicate. However, a dishonest adversary can have a chance to acquire some information that is prohibited to be leaked out. In the following paragraphs, we are going to discuss some possible attacks from the adversaries, especially from the database, that could affect security of the protocol.

\textit{\textbf{Data Privacy}}
As already discussed above, our protocol performs two rounds of communications to accomplish the querying task. However, a dishonest user can retrieve at most one item. This is guaranteed by the fact that only one (answering) qubit contains information on the item values and is transmitted exactly once during the protocol.
Indeed, by the Holevo bound \cite{NL00}, at most one bit of classical information can be retrieved from one qubit, in general. Therefore, the user will get exactly $A_j$ if he/she follows the protocol honestly. A dishonest user, however, may get information about other items but, no more than one bit of information on the data items in total. In comparison, with one extra round of communication designed specifically for the honesty detection, the QPQ protocol could leak at most two items/bits to a dishonest user.

If some small probability of error is accepted, a dishonest user may try to get some information about more data items. Indeed, using a special quantum interrogation procedure from \cite{Van98}, Theorem \ref{Thm_dprivacy} presented in the Appendix \ref{App_estimation} demonstrates that, given the initial state $|\psi_0\rangle$, it is possible to get some nontrivial information about $N/2$ items of the database in total. The same amount of information can be obtained when random guesses are used. Therefore, data privacy is well preserved in our protocol. It is worth to observe that no similar claims were given so far either when analyzing the QPQ protocol \cite{GLM08} or the phase-encoded private query protocols \cite{Ole12,Yu14}. In the Appendix \ref{App_estimation}, we show that given the initial state $\frac{1}{\sqrt{2}}(|0\rangle+|j\rangle)$ that was used both in the QPQ protocol \cite{GLM08}, and in one phase-encoded private query protocol \cite{Ole12}, an extra half of a bit of information on the database can be estimated, regardless of those obtained from a random guess, using the quantum interrogation procedure \cite{Van98}.

\textit{\textbf{User Privacy}}
In any attempt to extract information about $j$, which is the user privacy, the database may try to distinguish different choices of $j$ among all possible forms of the initial state $|\psi_0\rangle$. However, it can not obtain much information via this approach because there are $N$ potential states with a mutual overlap $1/\sqrt{N-1}+(N-2)/2(N-1)$ \footnote{For any two different choices of such potential states, for example, $|\varphi_0^i\rangle=\frac{1}{\sqrt{2}}\left(|j_i\rangle+\frac{1}{\sqrt{N-1}} \sum_{j'_i\neq j_i}|j'_1\rangle\right)$, where $i=1,2$, their overlap is defined to be $\langle\varphi_0^1|\varphi_0^2\rangle=1/\sqrt{N-1}+(N-2)/2(N-1)$.} that is close to $1/2$ for large $N$. A better approach for a dishonest database to learn the query is to apply a measurement directly on the received state, say, using the computational basis measurement ($\{|i\rangle\langle i|, i\in [N]\}$). This approach is easy-to-implement and may lead to a significant loss of information about the user privacy. The loss may be even $\log N$ bits in the QPQ protocol case and half of $\log N$ bits in the phase-encoded private query protocols cases. To make it even worse, the database in the phase-encoded private query protocols can conceal itself easily by merely sending back the outcome states $|j\rangle$ or $|0\rangle$.

As far as our new protocol is concerned, using the same approach to measure the received state in the computational basis, the cheating database may have outcomes distributed from $1$ to $N$, each with probability $1/\sqrt{2(N-1)}$, except for $j$ - in this case the probability is one-half. Due to the lack of knowledge about which index is the true query $j$ and which indexes are serving as rhetoric queries, the database can no longer simply identify a nonzero outcome (corresponding to an item index) as the true query $j$, even if $j$ has been measured. However, the database can make use of the second round of communication to tell whether or not it has indeed measured $j$. We explain the method as follows. Suppose that it has a measurement outcome $k$ at the first round of communication, the database can then use this $k$ and any index $l$ (different from $k$) to produce a fake state
\begin{equation*}
\frac{|k\rangle|1\rangle+|l\rangle|0\rangle}{\sqrt{2}}, \label{Eq_fakeforj}
\end{equation*}
\noindent and returns it to the user. Consequently, if the user keeps to follow the agreed steps, the user will return the state $\frac{1}{\sqrt{2}}(-|k\rangle|1\rangle+|l\rangle|0\rangle)$ if $k=j$, and $\frac{1}{\sqrt{2}}(|k\rangle|1\rangle+|l\rangle|0\rangle)$ otherwise. These two states are orthogonal and therefore the database can perform a discriminating measurement on the received state to determine whether or not $k=j$. Using such a strategy, the database can get $\frac{1}{2}\log N$ bits of information on $j$ during the protocol execution. Note that the information that the database has obtained on $j$ is $\frac{1}{2}\log N$ bits before and remains the same after the discriminating measurement is applied. What the database can gain via such a discriminating measurement is actually a bit of information on whether or not $k=j$.

The question is now, which state ought to be returned at the second round of communication in order to conceal the cheat? The database would most probably choose to return those states which can conceal its cheat successfully in any circumstance. Such a state, in the phase-encoded private query protocols \cite{Ole12, Yu14}, is the measurement outcome state ($|j\rangle$ or $|0\rangle$). In the basic form of our proposed protocol, it is the uniform state $\frac{1}{\sqrt{N}}\sum_{i=0}^{N-1}|i\rangle$ because such a state will not result in unexpected outcomes in user's measurement, i.e. when the user applies his/her measurement on such a state, the measurement outcome state can only be either $|\psi_3^+\rangle$ or $|\psi_3^-\rangle$. (We have that $1-|\langle\varphi|\psi_3^+\rangle|^2-|\langle\varphi|\psi_3^-\rangle|^2$ equals $0$, by replacing into it both $|\varphi\rangle=\frac{1}{\sqrt{N}}\sum_{i=0}^{N-1}|i\rangle$ and $|\psi_3^\pm\rangle=\frac{1}{\sqrt{2}}(\pm|j\rangle+\frac{1}{\sqrt{N-1}}\sum_{j'\neq j}|j'\rangle)$.) Since it can be created independently of the measurement outcomes of the database, the uniform state can, and only it can \footnote{The proof is omitted here. It can be done using ideas contained in the proof of Theorem \ref{Thm_cheatsens}.} help the database to evade the user's detection perfectly in the basic form of our proposed protocol. Furthermore, returning such a fake state will result in a nearly random answer being extracted by the user, as that by returning the measurement outcome state ($|j\rangle$ or $|0\rangle$) in the phase-encoded private query protocols.

\begin{table}[t]
\setlength{\abovecaptionskip}{0.cm}
\setlength{\belowcaptionskip}{-0.cm}
\tabcolsep 0pt
\caption{Security analysis results of the proposed protocol in comparison with  private query protocols of $\mathcal{O}(\log N)$ communication complexity.}
\vspace*{-12pt}
\begin{center}
\def\temptablewidth{1\textwidth}
{\rule{\temptablewidth}{1pt}}
\begin{tabular*}{\temptablewidth}{@{\extracolsep{\fill}}ccccc}
\multirow{2}*{} & \multicolumn{2}{c}{\tabincell{c}{\textbf{Data Privacy} \\ \textbf{(Leakage of $\bm{A}$)}}}  & \multicolumn{2}{c}{\textbf{User Privacy}}  \\   \cline{2-5}
            & \tabincell{c}{Deterministic} & \tabincell{c}{Nondeterministic \\ (via QIP)} & \tabincell{c}{Cheat- \\sensitive} & \tabincell{c}{b} \\   \hline
  \textbf{QPQ prot.}           & $2$ bits  & $1/2$ bit            & yes & $\log N$ bits   \\
  \textbf{Phase-enc. prot.} & $1$ bit & $1/2$ bit or not available            & no  & $\frac{1}{2}\log N$ bits   \\
  \textbf{Our new prot.}          & $1$ bit & $0$ bit & \ \ no$^{\rm a}$  & $\frac{1}{2}\log N$ bits
       \end{tabular*}
       {\rule{\temptablewidth}{1pt}}
       \footnotesize\raggedright{$^{\rm a}$ But it can be extended to a randomized form which is cheat-sensitive.} \\
       \footnotesize{$^{\rm b}$ Maximum Leakage of $j$ when cheating via a computational basis measurement.}
       \end{center}
       \label{tab:comparison}
       \end{table}

Security analysis results of our new protocol, as well as its comparison with both the QPQ protocol and the phase-encoded private query protocols discussed above, are summarized in the Table
\ref{tab:comparison}. Data presented in that table show the difference between the known communication-efficient private query protocols and our protocol concerning preservation of data privacy, user privacy and the cheat-sensitivity. Data in the column 2 in Table \ref{tab:comparison} tell that a dishonest database can learn deterministically at most one item in both protocols - ours and the phase-encoded private query protocols, and two items in the QPQ protocol. Data in the column 3 implies that given the initial state $|\psi_0\rangle$ in our protocol, a dishonest database can estimate, even if a small probability of an error is allowed, no more items than those obtained from a random guess. Such an analysis has been given neither for the QPQ protocol nor for the phase-encoded private query protocols. However, a proof sketch, given in the Appendix \ref{App_estimation} below, indicates that an extra half of a bit of information can be estimated by a dishonest user both in the QPQ protocol and in one of the phase-encoded private query protocols using quantum interrogation procedure, regardless of those bits obtained from a random guess.

Data in the column 4 of the table show that the QPQ protocol is cheat-sensitive while both phase-encoded private query protocols and the basic form of our protocol are not. But our protocol is extendable to be cheat-sensitive. We will present and explore one randomized form of our protocol in the next section along with a theorem showing its cheat-sensitivity. We have compared possible maximum leakages of user privacy among protocols in the column 5 in Table \ref{tab:comparison} when the database performs a computational basis measurement. There one can see that $\frac{1}{2}\log N$ bits of information on $j$ could leak out in both protocols, ours and the phase-encoded private query protocols. In the QPQ protocol, the entire $\log N$ bits of information on $j$ would leak out inevitably.

\section{Randomization of Our New Protocol}\label{Sec_variant}
Let us now consider the case that the user randomly selected $t$ ($1\leq t \leq N-1$) rhetoric queries different from the true query $j$, which constitute a rhetoric query set $T$, and use them to prepare the initial state, i.e. $|\psi'_0\rangle=\frac{1}{\sqrt{2}}\left(|j\rangle+\frac{\sum_{j'\in T}|j'\rangle}{\sqrt{t}}\right)$. As in the basic form of our protocol presented above, the new (randomized) protocol proceeds with the state $|\psi'_0\rangle$ to return the correct answer to the user \footnote{One of the two states $|\psi_3^{'\pm}\rangle=\frac{1}{\sqrt{2}}\left(\frac{\sum_{j'\in T}|j'\rangle}{\sqrt{t}}\pm|j\rangle\right)$ is then supposed to be returned to the user.}. However, the database can no longer evade detection by simply returning the uniform state $\frac{1}{\sqrt{N}}\sum_{i=0}^{N-1}|i\rangle$ because that would make it to be detected with a probability $1-|\langle\varphi|\psi_3^{'+}\rangle|^2-|\langle\varphi|\psi_3^{'-}\rangle|^2$, which, by replacing into it both $|\varphi\rangle=\frac{1}{\sqrt{N}}\sum_{i=0}^{N-1}|i\rangle$ and $|\psi_3^{'\pm}\rangle=\frac{1}{\sqrt{2}}\left(\frac{\sum_{j'\in T}|j'\rangle}{\sqrt{t}}\pm|j\rangle\right)$, equals $1-\frac{t+1}{N}$. Unless $t=N-1$, such a probability is nonzero. Now it is quite natural to ask whether there is any other state that can help the database to conceal its cheat in any circumstance in this randomized form of our protocol? If no such state can be found, then, we would say that the protocol is cheat-sensitive, because whichever state the database chooses to return, the user would have a non-zero probability to detect the cheat.

In our randomized protocol, the database is assumed to perform an arbitrary projective measurement on $|\psi'_0\rangle$ at the first round of communications. Through performing such measurements on $|\psi'_0\rangle$ database can gain some information on user privacy, $j$, and can then, in the next step, create a fake state and return it to the user in order to explore how much information it has actually obtained on $j$. At the second round of communication, the database creates another fake state and return it to the user for the purpose of concealing the cheat.

Now, we can prove the cheat-sensitivity of our protocol.
\begin{theorem}\label{Thm_cheatsens}
The randomized form of our new protocol is cheat-sensitive.
\end{theorem}

\proof{Suppose that the database has performed a projective measurement on the state $|\psi'_0\rangle$ at the first round of communication for the purpose of stealing information and has returned a fake state $|\varphi\rangle$ at the second round of communication for the purpose of concealing the cheat. In order to derive a cheat-sensitive property for the randomized form of our protocol, regardless of the measurement basis being used by the database at the first round, $|\varphi\rangle$ is assumed to be with a general form $\sum_{i=0}^{N-1}\alpha_i|i\rangle$ ($\sum_{i=0}^{N-1}|\alpha_i|^2=1$). The user can then apply a discriminating measurement (with two basis states $|\psi_3^{'\pm}\rangle$) on $|\varphi\rangle$. Probabilities of user's outcomes are $p^+=|\langle\varphi|\psi_3^{'+}\rangle|^2$ for $|\psi_3^{'+}\rangle$, $p^-=|\langle\varphi|\psi_3^{'-}\rangle|^2$ for $|\psi_3^{'-}\rangle$, and $p=1-p^+-p^-$ for neither of $|\psi_3^{'\pm}\rangle$, respectively. Then, the probability that the database is detected as cheating can be calculated as
\begin{small}
\begin{eqnarray}
p &=&  1-|\frac{1}{\sqrt{2t}}\sum\limits_{j'\in T}\alpha_{j'}+\frac{1}{\sqrt{2}}\alpha_j|^2-|\frac{1}{\sqrt{2t}}\sum\limits_{j'\in T}\alpha_{j'}-\frac{1}{\sqrt{2}}\alpha_j|^2  \nonumber \\
&=& 1-\frac{1}{t}|\sum\limits_{j'\in T}\alpha_{j'}|^2-|\alpha_j|^2  \nonumber \\
&\geq & 1-\sum\limits_{j'\in T}|\alpha_{j'}|^2-|\alpha_j|^2 \label{Eq_ProbinRandLw1}\\
&\geq & 1-\sum\limits_{i=0}^{N-1}|\alpha_i|^2=0 \label{Eq_ProbinRandLw2}
\end{eqnarray}
\end{small}

\noindent The inequality (\ref{Eq_ProbinRandLw1}) has been derived using the Cauchy-Schwarz inequality \footnote {The Cauchy-Schwarz inequality states that for complex numbers $\alpha_1,\cdots,\alpha_n$, $\beta_1,\cdots, \beta_n$,  $|\alpha_1\bar{\beta}_1+\cdots+\alpha_n\bar{\beta}_n|^2\leq(|\alpha_1|^2+\cdots+|\alpha_n|^2)(|\beta_1|^2+\cdots+|\beta_n|^2)$ (where the bar notation is used for complex conjugation). By setting $\beta_i=\frac{1}{\sqrt{n}}\ (1\leq i \leq n)$ we obtain that $\frac{1}{n}|\alpha_1+\cdots+\alpha_n|^2\leq |\alpha_1|^2+\cdots+|\alpha_n|^2$. } and the equality holds if and only if all $\alpha_{j'}$ ($j'\in T$) are equal. That equality in (\ref{Eq_ProbinRandLw2}) requires that all $\alpha_{i}$ ($i\neq j,j'$) have a zero value, which means that if it is not a special case in the basic form of our protocol that $t=N-1$, the database must know precisely both the true query $j$ and rhetoric queries $j'$ in $T$ for every query in order to create a fake state satisfying the equality in (\ref{Eq_ProbinRandLw2}). Otherwise, the database would be detected as cheating with a nonzero probability. However, that is impossible because a real query is encoded as a superposed state which is made from both the true query $j$ and a randomly selected nonempty set $T$ of rhetoric queries, and the database can never learn full knowledge on such a state simply through applying projective measurements on it.}

Actually, due to the lack of knowledge about both $j$ and $T$, the database can hardly recover the initial state $|\psi'_0\rangle$, after performing projective measurements on it. In particular, assuming that the database has obtained $k$ as an outcome of a computational basis measurement, the probability of recovering $|\psi'_0\rangle$ correctly is:
\begin{small}
\begin{eqnarray*}
q&=&q_{k=j}\times\frac{1}{\sum_{t=1}^{N-1}\binom{N-1}{t}}+q_{k\neq j}\times\frac{1}{\frac{1}{N-1}\times\sum_{t=1}^{N-1}\binom{N-2}{t-1}} \\
 &=&\frac{1}{2\cdot(2^{N-1}-1)}+\frac{1}{(N-1)\cdot2^{N-1}},
\end{eqnarray*}
\end{small}
\noindent where $q_{k=j}$ and $q_{k\neq j}$ are probabilities that the database has measured or has not measured $j$, respectively. Both of them equal to $\frac{1}{2}$. It is now easy to see that $q$ is close to 0 for large N.

As the next, it is natural to ask, which state is useful for the database to conceal its cheat, i.e. without knowing $T$, which state ought to be returned by the database to minimize the probability of its cheating being detected? For the sake of simplicity of our discussions, we consider only the case of the computational basis measurement. To deal with this question we have the following theorem.

\begin{theorem}\label{optimal}
Let $|\psi'_0\rangle=\frac{1}{\sqrt{2}}\left(|j\rangle+\frac{\sum_{j'\in T}|j'\rangle}{\sqrt{t}}\right)(1\leq t \leq N-1)$ be the initial state of our new protocol without $T$ being known to the database. Suppose that a dishonest database has performed a computational basis measurement and has obtained $j$. Then the database will return the state $|j\rangle$. Suppose that it has obtained a $k$ different from $j$, the database would most probably return the fake state $|\varphi\rangle=\frac{2}{\sqrt{N+3}}|k\rangle+\sum_{k'\neq k}\frac{1}{\sqrt{N+3}}|k'\rangle$.
\end{theorem}

\proof{Firstly, it is safe for the database, if $j$ has been obtained by a computational basis measurement, to return the measurement outcome state $|j\rangle$ in order to conceal its cheat. The probability that the database is detected cheating is $0$ in such a case. Secondly, in case that $j$ has not been obtained by a computational basis measurement, but some other $k$, what the database can do to minimize the probability of being detected in user's measurement is, according to the inequality (\ref{Eq_ProbinRandLw1}), to create a fake state $|\varphi\rangle$ of the form $a|k\rangle+\sum_{k'\neq k}b|k'\rangle$, where, without loss of generality, $a$ and $b$ are real numbers satisfying the equality $a^2+(N-1)b^2=1$. Then, for each $t$, $a$, and $b$, the probability of the database being detected is
\begin{small}
\begin{eqnarray}
p_{t,a,b}&=& 1-\frac{\left(a+b(t-1)\right)^2}{t}-b^2 \nonumber\\
 &=& 1-2ab+b^2-b^2t-\frac{a^2+b^2-2ab}{t}. \label{Eq_PinR_tab}
\end{eqnarray}
\end{small}
\noindent Then, the expected values of $p_{t,a,b}$ for all $t$ is $\bar{p}_{a,b}=\sum_{t=1}^{N-1}\binom{N-1}{t}\cdot p_{t,a,b}/(2^{N-1}-1)$, where $2^{N-1}-1$ is the number of candidate initial states. Using the equalities $\sum_{t=1}^{N-1}t\cdot\binom{N-1}{t}=(N-1)\times 2^{N-2}$ and $\sum_{t=1}^{N-1}\binom{N-1}{t}= 2^{N-1}-1$, we get that
\begin{small}
\begin{eqnarray*}
\bar{p}_{a,b}&\approx& 1-2ab-\frac{N-3}{2}\cdot b^2-\frac{a^2+b^2-2ab}{2^{N-1}-1}\cdot\sum_{t=1}^{N-1}\binom{N-1}{t}\frac{1}{t}\ \ .
\end{eqnarray*}
\end{small}

\noindent Using one equality from \cite{Tb65}, we have for $N\geq 2$, $\sum_{t=1}^{N-1}\frac{1}{t}\cdot \binom{N-1}{t}\leq \sum_{t=1}^{N-1}\frac{2}{t+1}\cdot \binom{N-1}{t}=2\cdot(\frac{2^N-1}{N}-1),$ We therefore have for $N\geq 2$,
\begin{small}
\begin{eqnarray*}
\bar{p}_{a,b}&\geq& 1-2ab-\frac{N-3}{2}\cdot b^2-(a^2+b^2-2ab)\cdot\frac{4}{N}.
\end{eqnarray*}
\end{small}

\noindent By using  the equality $a^2+(N-1)b^2=1$, we get $\bar{p}_{a,b}= \frac{1}{2}+\frac{1}{2}(a-2b)^2-b^2-(a-b)^2\cdot\frac{4}{N}$, which has a minimum value approximately $\frac{1}{2}-\frac{1}{N+3}$ at $a=\frac{2}{\sqrt{N+3}}$ and $b=\frac{1}{\sqrt{N+3}}$. That means that in case of outcomes different from $j$, returning $|\varphi\rangle=\frac{2}{\sqrt{N+3}}|k\rangle+\sum_{k'\neq k}\frac{1}{\sqrt{N+3}}|k'\rangle$ can minimize the probability of the database being revealed to approximately $\frac{1}{2}$. Therefore, the overall probability that the database has a minimum chance to be detected cheating is approximately $\frac{1}{2}\times 0+\frac{1}{2}\times\frac{1}{2}=\frac{1}{4}$. }

The next natural question is: how many rhetoric queries are optimal for the user to reveal a database cheating via the computational basis measurement, i.e. which value of $t$ will maximize the probability of detecting such a cheat? We start with the following theorem:

\begin{theorem}\label{Thm_optimal}
$t=\sqrt{N-\frac{4}{\pi}\sqrt{N-1}}$ is the optimal number of rhetoric queries for the user to be defended against a cheating via the computational basis measurement.
\end{theorem}

\proof{Suppose that a dishonest database has performed a computational basis measurement and obtained $k$ as an outcome. Then, in case that $k=j$, no matter what the value of $t$ is, the database can return $|j\rangle$ at no risk of being detected as cheating. In other cases, namely when $k\neq j$, according to the inequality (\ref{Eq_ProbinRandLw1}), the database would return a fake state of the form $|\varphi\rangle=a|k\rangle+\sum_{k'\neq k}b|k'\rangle$ with $a^2+(N-1)b^2=1$ in order to minimize the probability of being detected in user's measurement. We rewrite $p_{t,a,b}$ in the equation (\ref{Eq_PinR_tab}) to $p_{t,\alpha}=1-\frac{\cos^2\alpha}{t}-\frac{\sin^2\alpha}{N-1}(\frac{1}{t}-1+t)-\frac{2\sin\alpha\cos\alpha}{\sqrt{N-1}}(1-\frac{1}{t})$, by setting that $a=\cos\alpha$ and $b=\frac{\sin\alpha}{\sqrt{N-1}}$ ($0\leq \alpha \leq \frac{\pi}{2}$). Then, the expected values of $p_{t,\alpha}$ for all $\alpha$ is
\begin{small}
\begin{eqnarray}
\bar{p}_t\times\frac{\pi}{2}&=& \int_0^{\frac{\pi}{2}}p_{t,\alpha}\mathbbm{d}\mathbbm{\alpha} = \frac{\pi}{2}-(\frac{\alpha}{2}+\frac{\sin2\alpha}{4})\big{|}_{0}^{\frac{\pi}{2}}\times\frac{1}{t}- \nonumber\\
&& \left((\frac{\alpha}{2}-\frac{\sin2\alpha}{4})\big{|}_{0}^{\frac{\pi}{2}}\cdot\frac{1}{N-1}\cdot(\frac{1}{t}+t-1)\right)-\frac{\sin^2\alpha}{\sqrt{N-1}}\big{|}_{0}^{\frac{\pi}{2}}\times(1-\frac{1}{t}) \nonumber \\
&=& \frac{(2N-1)\pi}{4(N-1)}-\frac{1}{\sqrt{N-1}}-\left(\left(\frac{N\pi}{4(N-1)}-\frac{1}{\sqrt{N-1}}\right)\cdot\frac{1}{t}+\frac{\pi}{4(N-1)}\cdot t\right) \nonumber\\
&\leq & \frac{(2N-1)\pi}{4(N-1)}-\frac{1}{\sqrt{N-1}}-\frac{\pi}{2(N-1)}\sqrt{N-\frac{4}{\pi}\sqrt{N-1}}. \label{Eq_opt_last}
\end{eqnarray}
\end{small}

\noindent The equality in (\ref{Eq_opt_last}) holds if and only if $t=\sqrt{N-\frac{4}{\pi}\sqrt{N-1}}$. Therefore, $\sqrt{N-\frac{4}{\pi}\sqrt{N-1}}$ rhetoric queries are the optimal number of rhetoric queries for the user to reveal such a cheating activity in the randomized variant of our protocol.}

\begin{corollary}\label{Col_optimal}
The maximum of probability that the user will detect a computational basis measurement cheating activity is $p_{maxi}=\frac{1}{2}-\frac{1}{\pi\sqrt{N-1}}-\frac{2\sqrt{N-\frac{4}{\pi}\sqrt{N-1}}-1}{4(N-1)}$.
\end{corollary}

\proof{The user will detect a cheat with a zero probability when the database has obtained $j$ from a computational basis measurement, and by Theorem \ref{Thm_optimal}, with a maximum of probability $\bar{p}=1-\frac{2}{\pi\sqrt{N-1}}-\frac{2\sqrt{N-\frac{4}{\pi}\sqrt{N-1}}-1}{2(N-1)}$ at $t=\sqrt{N-\frac{4}{\pi}\sqrt{N-1}}$ when the database has not obtained $j$. Therefore, the maximum of probability that the user will detect a dishonest database cheating via a computational basis measurement is $p_{maxi}=\frac{1}{2}\times 0+\frac{1}{2}\times \bar{p}=\frac{1}{2}-\frac{1}{\pi\sqrt{N-1}}-\frac{2\sqrt{N-\frac{4}{\pi}\sqrt{N-1}}-1}{4(N-1)}$ by using $\sqrt{N-\frac{4}{\pi}\sqrt{N-1}}$ rhetoric queries.
}

The advantage of using an (flexible) initial state $|\psi'_0\rangle$ over (the fixed) $|\psi_0\rangle$ in preserving user privacy comes mainly from those cases that the database has not obtained $j$ from the measurement. In such cases, the database can recover the initial state only when guessing correctly both $j$ and all rhetorical queries.

$p_{maxi}$ is approximately $\frac{1}{2}$ for large $N$. In the basic form of the protocol, the database performing projective measurements would evade detection by returning the uniform state, regardless of the measurement basis being used. Therefore, the user privacy is significantly enhanced by using randomness on rhetoric queries in our new protocol.

\section{Conclusions}\label{Sec_conclusion}
In this paper we have described and analyzed a new quantum private database query protocol which is a generalization of the known communication-efficient protocols, namely, the QPQ protocol and the phase-encoded private query protocols. The proposed protocol has more than one item (index) to serve as rhetoric queries, with no need for the user to know answers to such rhetoric queries beforehand in order to guarantee a successful query. The database answers queries via a special data retrieving algorithm. The user can retrieve the answer $A_j$ to the query $j$ through an encoding-decoding process.

By exchanging only $\mathcal{O}(\log N)$ of data to accomplish a query task, our new protocol maintains the advantage of the current protocols \cite{GLM08,Ole12,Yu14} for the same task with respect to communication complexity. It also achieves an improvement in security, especially in a protection of the user privacy. Comparing with the QPQ protocol \cite{GLM08}, in which the entire amount of user privacy would inevitably leak to a dishonest database performing a computational basis measurement, at most half amount of user privacy could leak out with respect to the same attack when our new protocol is used. Comparing with the phase-encoded private query protocols, one randomized form of our new protocol has been proved cheat-sensitive. A dishonest user would be able to retrieve at most one item deterministically, which is the same as in the phase-encoded private query protocols, but one less than in the QPQ protocol. It is also proved in the Appendix \ref{App_estimation} that even if a small probability of an error is allowed, no one can estimate extra item, via the quantum interrogation procedure, than that can be obtained from a random guess. A similar evaluation has not been yet provided either for the QPQ protocol or for phase-encoded private query protocols. It is shown in Appendix \ref{App_estimation} that, by using a quantum interrogation procedure, an extra half of a bit of information on database can be estimated for both protocols, regardless of those obtained from a random guess.

In the basic form of the proposed protocol, all item indexes but $j$ serve as rhetoric queries. A dishonest database performing projective measurements can return a fake uniform state to evade detection, regardless of measurement basis being used. By introducing randomness on rhetoric queries, our protocol gains a cheat-sensitive property, as demonstrated by Theorem \ref{Thm_cheatsens}. In the mean while, the user will reveal a database performing a computational basis measurement with a maximum probability approximately $\frac{1}{2}$ by using $\mathcal{O}(\sqrt{N})$ rhetoric queries for large $N$, which are demonstrated by Theorem \ref{Thm_optimal} and Corollary \ref{Col_optimal}.

\begin{appendices}
\section{Estimation of a Leakage of Data Privacy}\label{App_estimation}
The user may estimate a number of database items, if a small probability of error is allowed, by using a quantum interrogation procedure from Section 4.2 in \cite{Van98}, given the initial state $|\psi_0\rangle$. Indeed, we can prove the following theorem:

\begin{theorem}\label{Thm_dprivacy}
Let $|\psi_0\rangle=\frac{1}{\sqrt{2}}\left(|j\rangle+\frac{\sum_{j'\neq j}|j'\rangle}{\sqrt{N-1}}\right)$ be the initial state of the above mentioned quantum interrogation procedure. A user can use this procedure to estimate no more than $\frac{N}{2}$ expected number of database items, which is equal to that obtained using random guesses.
\end{theorem}

\proof{
By attaching an appropriate number of ancillary qubits, the state $|\psi_0\rangle$ can be mapped into the state
$|\psi'_0\rangle=\frac{1}{\sqrt{2}}(|\vec{x}_j\rangle+\frac{1}{\sqrt{N-1}}\sum_{j'\neq j}|\vec{x}_{j'}\rangle)$, where $\vec{x_i}\in \{0,1\}^N$ has  a $1$ at the $i$th position. The proof is based on the method from the page 11 in \cite{Van98}. As the first step, using one auxiliary qubit in the state $(|0\rangle-|1\rangle)/\sqrt{2}$ attached, a special data retrieving algorithm or an oracle $A$ maps $|\psi'_0\rangle$ into the state

\begin{equation*}
A|\psi'_0\rangle=\frac{1}{\sqrt{2}}\cdot(-1)^{\vec{x}_j\cdot\vec{A}}|\vec{x}_j\rangle+\frac{1}{\sqrt{2(N-1)}}\sum\limits_{j'\neq j}(-1)^{\vec{x}_{j'}\cdot\vec{A}}|\vec{x}_{j'}\rangle,
\end{equation*}

\noindent where $\vec{A}$ denotes an unknown database string. By assuming that $\vec{A}$ consists of zeros only, the initial state $|\psi'_0\rangle$ will not change after being applied by the oracle $A$. Therefore, we have:

\begin{small}
\begin{eqnarray*}
H^{\otimes N}A|\psi'_0\rangle &=& H^{\otimes N} \left(\frac{1}{\sqrt{2}}|\vec{x}_j\rangle+\frac{1}{\sqrt{2(N-1)}}\sum_{j'\neq j}|\vec{x}_{j'}\rangle\right) \\
&=&  \frac{1}{\sqrt{2^N}}\sum\limits_{\vec{y}}\left(\frac{1}{\sqrt{2}}\cdot (-1)^{\vec{x}_j\cdot\vec{y}}+\frac{1}{\sqrt{2(N-1)}}\sum\limits_{j'\neq j}(-1)^{\vec{x}_{j'}\cdot\vec{y}}\right)|\vec{y}\rangle \\
&=&  \frac{1}{\sqrt{2^N}}\sum\limits_{\vec{y}}\left(\frac{(-1)^{y_j}}{\sqrt{2}}+\frac{2t_{\vec{y}}-N-(-1)^{y_j}}{\sqrt{2(N-1)}}\right)|\vec{y}\rangle \\
&=&  \frac{1}{\sqrt{2^N}}\left(\sum\limits_{y_j=0}\left(\frac{1}{\sqrt{2}}-\frac{N+1}{\sqrt{2(N-1)}}+\frac{\sqrt{2}t_{\vec{y}}}{\sqrt{N-1}}\right)|\vec{y}\rangle+\right.\\
&& \ \ \ \ \ \ \ \ \ \ \ \ \ \ \ \left.\sum\limits_{y_j=1}\left(-\frac{1}{\sqrt{2}}-\frac{\sqrt{N-1}}{\sqrt{2}}+\frac{\sqrt{2}t_{\vec{y}}}{\sqrt{N-1}}\right)|\vec{y}\rangle\right),
\end{eqnarray*}
\end{small}

\noindent where $y_i$ and $t_{\vec{y}}$ are the $i$th bit and the number of zeros of the $N$-bit binary string $\vec{y}$, respectively. Let us denote coefficients of those $|\vec{y}\rangle$ with $y_j=0$ and $y_j=1$ in the above last equation as $a_{t_{\vec{y}}}$ and $b_{t_{\vec{y}}}$, respectively. They depend solely on the number of zeros of $\vec{y}$. The expected number of correct bits of $\vec{A}$ equals therefore the expected number of zeros of the string $\vec{y}$, which is
\begin{equation*}
\#zeros(H^{\otimes N}A|\psi'_0\rangle)= \sum\limits_{t=0}^{N}\left(t\cdot\binom{N-1}{t-1}\cdot a_t^2+t\cdot\binom{N-1}{t}\cdot b_t^2\right),
\end{equation*}

\noindent where $H$ denotes the Hadamard operator. Using the following equalities
\begin{equation*}
\begin{array}{l@{\ }l}
\sum\limits_{t=1}^{N}t\binom{N-1}{t-1}= (N+1)\cdot 2^{N-2} & \sum\limits_{t=0}^{N-1}t\binom{N-1}{t}=(N-1)\cdot 2^{N-2}, \\
\sum\limits_{t=1}^{N}t^2\binom{N-1}{t-1}= N(N+3)\cdot 2^{N-3} & \sum\limits_{t=0}^{N-1}t^2\binom{N-1}{t} = N(N-1)\cdot 2^{N-3},\\
\sum\limits_{t=1}^{N}t^3\binom{N-1}{t-1}= (N+1)(N^2+5N-2)\cdot 2^{N-4} & \sum\limits_{t=0}^{N-1}t^3\binom{N-1}{t} =(N-1)^2(N+2)\cdot 2^{N-4},
\end{array}
\end{equation*}

\noindent a straightforward but cumbersome calculation shows that the quantity is $N/2$, which is equal to that obtained from a random guess.}

It is worth to note that with the initial state $\frac{1}{\sqrt{2}}|0\cdots0\rangle+\frac{1}{\sqrt{2N}}\sum_{i=0}^{N-1}|\vec{x}_i\rangle$, about $\frac{N}{2}+\frac{\sqrt{N}}{2}$ correct bits of $\vec{A}$ can be estimated in total, as shown in the Section 4.3 in \cite{Van98}. As next we show, by a similar proof, for the QPQ protocol \cite{GLM08} and also for one of the phase-encoded private query protocols \cite{Ole12} that with the initial state $|\phi\rangle=\frac{1}{\sqrt{2}}(|0\rangle+|j\rangle)$ (for which the transformed initial state is $|\phi'\rangle=\frac{1}{\sqrt{2}}(|\vec{0}\rangle+|\vec{x}_j\rangle)$ ), the expected number of correct bits of $\vec{A}$ is
\begin{eqnarray*}
\#zeros(H^{\otimes N}A|\phi'\rangle) &=& \#zeros\left(\frac{1}{\sqrt{2^N}}\sum\limits_{\vec{y}}\left(\frac{1}{\sqrt{2}}+\frac{(-1)^{y_j}}{\sqrt{2}}\right)|\vec{y}\rangle\right) \\
&=& \frac{1}{2^N}\cdot\sum\limits_{t=1}^{N}\left(t\cdot\binom{N-1}{t-1}\cdot (\sqrt{2})^2\right)=\frac{N}{2}+\frac{1}{2}.
\end{eqnarray*}

\end{appendices}

\section*{Dedication and Acknowledgments}

The authors would like to thank the anonymous reviewers' comments and suggestions that help to improve the quality of the manuscript.
The paper is dedicated to a great man of theoretical computer science, to Maurice Nivat, especially for his understanding that the field should be taken very broadly and steps should be taken to incorporate in its development an  enormous research potential of the communities also outside of Europe and North America.

This work was supported in part by the National Natural Science Foundation (Grant Nos. 61572532,  61876195,  61472452, 61772565), the Fundamental Research Funds for the Central Universities of China (Nos. 21617402, 83216003023),  the Joint Funds of the National Natural Science Foundation of China and China General Technology Research Institute (Grant No. U1736113), the Natural Science Foundation of Guangdong Province of China (Grant Nos. 2017B030311011, 2017A030313378). It has also been supported by the Faculty of Informatics of the Masaryk university in Brno, Czech Republic.

\end{document}